\documentclass[12pt]{article}

\usepackage{amssymb}
\usepackage{amsmath}
\usepackage{graphicx}
\usepackage{latexsym}

\begin{document}

\begin{titlepage}
\begin{center}

\hfill TU-790 \\
\hfill STUPP-07-191 \\
\hfill \today

\vspace{0.5cm}
{\large\bf Relic abundance of dark matter
in universal extra dimension models with right-handed neutrinos}
\vspace{1cm}

{\bf Shigeki Matsumoto}$^{a,\,}
$\footnote{smatsu@tuhep.phys.tohoku.ac.jp},
{\bf Joe Sato}$^{b,\,}
$\footnote{joe@phy.saitama-u.ac.jp},
{\bf Masato Senami}$^{c,\,}
$\footnote{senami@icrr.u-tokyo.ac.jp}, \\
and 
{\bf Masato Yamanaka}$^{b,\,}
$\footnote{masa@krishna.phy.saitama-u.ac.jp} \\

\vskip 0.15in

{\it
$^a${Institute for International Advanced Interdisciplinary Research,
     Tohoku University, Sendai, Miyagi 980-8578, Japan} \\
$^b${Department of Physics, Saitama University, 
     Shimo-okubo, Sakura-ku, Saitama, 338-8570, Japan} \\
$^c${ICRR, University of Tokyo, Kashiwa, Chiba 277-8582, Japan }
}

\vskip 0.4in

\abstract{
Relic abundance of dark matter is investigated in the framework of universal
extra dimension models with right-handed neutrinos. These models are free from
the serious Kaluza-Klein (KK) graviton problem that the original universal
extra dimension model has. The first KK particle of the right-handed neutrino
is a candidate for dark matter in this framework, and its relic abundance is
determined by three processes, (1) the decay of the KK photon into the first KK
right-handed neutrino in the late universe, (2) production of the first KK
right-handed neutrino from the thermal bath in the early universe, and (3) the
decay of higher KK right-handed neutrinos into the first KK right-handed
neutrino in the late universe. When ordinary neutrino masses are large enough
such as the degenerate mass spectrum case, the last process contribute to the
abundance significantly, even if the reheating temperature is low. The scale of
the extra dimension consistent with cosmological observations can be 500 GeV in
the minimal setup of universal extra dimension models with right-handed
neutrinos.}

\end{center}
\end{titlepage}
\setcounter{footnote}{0}

\section{Introduction}

The existence of dark matter has already been established by excellent
cosmological observations such as Wilkinson Microwave Anisotropy Probe
(WMAP) \cite{WMAP}. The constituent of dark matter is, however, still
unknown. There is no candidate for dark matter in the standard model
(SM). Hence, its existence leads us to the consideration for physics
beyond the SM. Many extensions of the SM including a dark matter
candidate have been proposed so far. Among those, a universal extra
dimension (UED) scenario \cite{Appelquist:2000nn},
which is motivated by TeV scale extra dimension theory \cite{Antoniadis},
is one of interesting extensions for new physics.
Because the lightest Kaluza-Klein (KK)
particle is stable due to the KK parity conservation which originates in
the momentum conservation along an extra dimension, and is a good
candidate for dark matter \cite{Servant:2002aq}.

However, the original UED model has the serious problem called the KK
graviton problem, when we take the KK graviton into account
\cite{Matsumoto:2005uh, Feng:2003xh}. In the parameter region
where the scale of the extra dimension is smaller than 800 GeV,
the KK graviton is
the lightest KK particle (LKP), and the next LKP (NLKP) is the KK
photon. Hence, the KK photon decays into a KK graviton and a photon with
very long lifetime because of interactions suppressed by the Planck
scale. This fact leads to the serious cosmological problem; KK photons
produced in the early universe decay into photons (and KK gravitons) in
the late universe, and these photons distort the cosmic microwave
background (CMB) spectrum or the diffuse photon spectrum
\cite{Feng:2003xh}.

Recently, it has been pointed out that the KK graviton problem is solved
by an introduction of right-handed neutrinos into the UED scenario
\cite{Matsumoto:2006bf}.  Another motivation of the introduction is to
describe neutrino masses in the scenario.
In this model, the mass of the KK right-handed neutrino
is smaller than that of the KK photon
if the mass of the zero mode of the right-handed neutrino is small enough.
Therefore, the KK photon decays
into a KK right-handed neutrino and an ordinary neutrino instead of a KK
graviton and a photon. Thus, the decay of the KK photon is not
accompanied by a photon in the leading process, and the diffuse photon
or CMB spectrum is not distorted. The introduction of right-handed
neutrinos remedies two shortcomings of the UED scenario simultaneously;
the KK graviton problem and the description of neutrino masses.

In this paper, we investigate the relic abundance of dark matter in the
framework of UED models with right-handed neutrinos.\footnote{
Some works about UED models with right-handed neutrinos
also exist in Ref.~\cite{Hsieh:2006fg}.}
The first KK
particle of the right-handed neutrino is dark matter in this setup. The
first KK right-handed neutrino never reaches thermal equilibrium in the
early universe due to the very weak Yukawa interaction. Nevertheless,
first KK right-handed neutrinos are gradually produced from the thermal
bath in the early universe, which contribute to the relic abundance of
dark matter as in the case of the gravitino production in supersymmetric
models. Furthermore, not only first KK right-handed neutrinos but also
higher KK right-handed neutrinos are also produced in the early
universe, and they produce first KK right-handed neutrinos through
cascade decay in the late universe. Due to the copious production of
higher KK right-handed neutrinos, the contribution to the abundance can
be significant, even if the reheating temperature is as low as 10
TeV. In addition to these processes, another process contributes to the
abundance, which is the nonthermal production from the decay of KK
photons in the late universe. As a result of these contributions to the
abundance, the scale of the extra dimension consistent with the WMAP
observation is significantly altered compared to UED models without
right-handed neutrinos.  The scale can be 500 GeV in the minimal setup
of the UED model with right-handed neutrinos.

This paper is organized as follows. First, we briefly summarize the
minimal UED (MUED) model and the extension with right-handed neutrinos
in the next section.  Thermal corrections to masses of KK particles,
which play an important role for the production of the KK right-handed
neutrinos in the early universe, are discussed in Section 3. In Section
4, the calculation procedure of the relic abundance of the dark matter is
shown. The numerical results are presented in Section 5.  Section 6 is
devoted to summary.

\section{MUED Model with right-handed neutrinos}

In order to estimate the relic abundance of dark matter quantitatively,
we consider the MUED model with right-handed neutrinos. We briefly
review the model in this section, particularly focusing on the production
of KK right-handed neutrinos in the early universe.

Among various models in the UED scenario, the simplest one is the MUED
model.  This model is the simplest extension of the SM in the
five-dimensional space-time, where the extra dimension is compactified
on an $S^1/Z_2$ orbifold with the radius $R$. The model can be described
by the SM particles and their KK modes from the four-dimensional view
point. The compactification by the orbifold is required to describe the
SM chiral fermions as KK zero modes, because a fermion is represented as
a four-component spinor in a five dimensional space-time.

The orbifold has two fixed points in the extra dimension, and its existence
breaks the translational invariance along the extra dimension. Hence, the KK
number conservation, which is nothing but the momentum conservation along the
extra dimension, is broken by radiative corrections. The subgroup in the KK
number conservation called KK parity, however, remains unbroken. Under this
parity, particles with even (odd) KK number have plus (minus) charge.
Therefore, the LKP is stable as mentioned in the previous section.

The SM particles and their KK partners have identical gauge charges. All
bulk interactions relevant to KK particles are determined by the SM
Lagrangian. On the other hand, boundary interactions are not directly
related to the SM interactions. In the MUED model, it is postulated that
these boundary interactions vanish at the cutoff scale $\Lambda$
\cite{Cheng:2002iz}, which should be $\Lambda R \sim {\cal O} (10)$
\cite{Appelquist:2000nn, Bhattacharyya:2006ym}, and
they appear only through renormalization group evolutions.
This means that the MUED
model should be regarded as an effective theory describing physics below
the cutoff scale $\Lambda$. Hence, the model is very restrictive and has
only two new physics parameters, $\Lambda$ and $1/R$. Furthermore, the
dependence of physical quantities on $\Lambda$ is very weak, because it
always appears with a loop suppression, and gives logarithmic
corrections. The MUED model gives definite predictions in spite of its
non-renormalizable nature.

The constraint to the model from collider experiments is not so stringent,
because first KK particles are necessarily
produced in a pair. The situation is similar to supersymmetric models with
R-parity. In fact, the value of $1/R$ is constrained by electroweak
precision measurements as $1/R \gtrsim$ 400 GeV depending on the Higgs
mass \cite{Appelquist:2000nn, Kakizaki:2006dz, Appelquist:2002wb, Gogoladze}. More
stringent bound is recently reported in Ref.~\cite{Haisch:2007vb}, which
comes from $b \to s \gamma$ process. It gives the constraint as $ 1/R >
$ 600 GeV at 95\% confidence level, which is independent of the Higgs mass.

Here, we introduce right-handed neutrinos into the MUED model in order to
describe the neutrino mass. In the five dimensional description, neutrino
masses are expressed by Dirac and Majorana mass terms as
\begin{eqnarray}
 {\cal L}_{\nu{\rm -mass}}
 = y_\nu \bar N L \Phi
 + M N^T C \gamma_5 N + {\rm h.c.}~,
 \label{Neutrino Yukawa}
\end{eqnarray}
where $N$ is the right-handed neutrino, $L$ is the left-handed (doublet)
lepton, and $C \equiv i \gamma_0 \gamma_2$ is the charge conjugation operator.
The flavor index is omitted here. According to the algebraic structure of the
five dimensional space-time, the ordinary Majorana mass term in the five
dimensional space-time, $N^T C N$, is not allowed
\cite{Weinberg:1984vb, Pilaftsis:1999jk}. Nevertheless, the second term gives
the Majorana mass for the right-handed neutrino after the compactification
\cite{Pilaftsis:1999jk}. 
Because UED models are an
effective theory below about 10 TeV, the right-handed neutrino mass
should be less than 10 TeV.
As a result, very small Yukawa interactions are inevitably introduced.
Hence, we will consider only the Dirac mass term and ignore the Majorana mass term for
definiteness\footnote{If we impose the lepton number symmetry, the absence of
Majorana masses is realized.}. Even if the Majorana mass term is included,
our discussion can be easily extended in a straightforward manner,
unless the right-handed neutrino is not thermalized in the early universe.

One of the specific features of UED models is the mass degeneracy
between KK particles with the same KK number $n$, where their masses are
approximately $n/R$.  Mass differences between these particles are
induced by radiative corrections. Since the corrections are small
compared with $n/R$, KK particles are still degenerate in mass. The LKP is
the first KK graviton and the NLKP is the first KK right-handed neutrino
in the broad parameter region of the MUED model with right-handed
neutrinos. However, the decay process of the KK right-handed neutrino
into the KK graviton is not allowed kinematically.  Thus, both the KK
particles can be dark matter. Next to NLKP is the first KK photon and it
decays dominantly into the KK right-handed neutrino and an ordinary
neutrino. Therefore, KK photons produced in the early universe decay
into KK right-handed neutrinos by emitting neutrinos in the late
universe, and the model is free from the dangerous KK graviton
problem\footnote{In addition to the KK graviton problem, there is
another problem called the radion problem, which is the common problem
in extra dimension scenarios. As this problem is not directly related to
our discussion, we ignore this problem. For the detailed discussion of
the problem, see Ref.~\cite{Kolb:2003mm}.}.

In the early universe, KK right-handed neutrinos are also produced from
the thermal bath. Since the right-handed neutrino is gauge singlet, it
has only a Yukawa interaction with the coupling $y_\nu = m_\nu / v$,
where $m_\nu$ is a neutrino mass and $v \sim 246$ GeV is the vacuum
expectation value of the Higgs field.  Hence, the production of KK
right-handed neutrinos originates only in the Yukawa interaction.

Before going to the discussion of the production process, we summarize KK
particles relevant to the interaction. First, we discuss KK leptons. The mass
matrix of KK leptons are
\begin{eqnarray}
  \left(
  \begin{array}{cc}
   n / R + \delta m_{l^{(n)}} & m_l
   \\
   m_l & - n / R - \delta m_{L^{(n)}}
  \end{array}
 \right)~,
 \label{eq:KKleptonmass}
\end{eqnarray}
where $l(L)$ represents the SU(2) doublet (singlet) lepton and $m_l$ is the
corresponding SM lepton mass, which vanishes in the early universe before the
electroweak symmetry breaking. When the off-diagonal elements are absent, the
mass eigenstates of KK leptons are determined by the diagonal elements.
Radiative corrections to the right-handed neutrino are negligibly small due to
the smallness of $y_\nu$ as $\delta m_{N^{(n)}} \sim (n/R) y_\nu^2 / (16\pi^2)
\ln (\Lambda^2 R^2 / n^2)$. Radiative corrections for doublet leptons,
$\delta m_{l^{(n)}}$, are given in Ref.~\cite{Cheng:2002iz}.

Next, KK Higgs bosons are discussed. Since the KK modes of the Higgs
field are not absorbed in KK gauge bosons, all of them, neutral scalar
$H^{(n)}$, pseudo scalar $A^{(n)}$ and charged scalars $H^{\pm(n)}$,
remain as physical states.  The latter three are KK modes of the
Nambu-Goldstone particles in the SM. Then, the KK Higgs doublet
$\Phi^{(n)} $ is expressed as $\Phi^{(n)} = ( H^{+(n)} , (H^{(n)} + i
A^{(n)}) / \sqrt{2} )$. The masses of the KK Higgs bosons are given in
Ref.~\cite{Cheng:2002iz}. We use the notation $m_{\Phi^{(n)}}$ for the
mass of the Higgs bosons with KK number $n$ collectively,
because they have the same
mass before the electroweak symmetry breaking.

Here, let us consider production processes of the KK right-handed
neutrinos in the early universe.  Production by some particle decay
dominate over that by scattering in the thermal bath as in the case of
right-handed sneutrino productions in supersymmetric models
\cite{Asaka:2005cn}. In our model, the process occurs through the Yukawa
interaction\footnote{We do not have to worry about other decay processes
such as $W^{(l)} \to \bar N^{(m)} + L^{(n)}$ in the early universe,
because these decays occur through the chirality flip via the Dirac
neutrino mass term, which does not exist before the electroweak symmetry
breaking.} in Eq.~(\ref{Neutrino Yukawa}); $\Phi \rightarrow \bar{L} N
(\Phi^* \to L \bar{N}$) or $L \rightarrow \Phi N (\bar{L} \rightarrow \Phi^*
\bar{N})$. At first sight, one might wonder if these processes are
highly suppressed by tiny phase space due to the mass degeneracy of KK
particles and the KK number conservation at tree level
diagrams. However, this is not true in the early universe, because the
masses of the KK Higgs particles are drastically increased by thermal
corrections. We discuss this point in the next section. Thus, the
dominant production process of KK right-handed neutrinos in the early
universe is the decay of Higgs bosons. In Section 4, we will investigate that
how large contribution to the relic abundance of dark matter comes
through the process. In addition to that, the late-time decay of the KK
photon also contributes to the abundance. We also discuss this process
in Section 4.

\section{Thermal masses}

It is well known that the mass of a particle receives a correction by thermal effects, 
when the particle is immersed in the thermal bath.
In particular, the temperature of the bath is high enough,
then the correction becomes large.
For example, the Higgs boson in the SM receives the following corrections
at the high temperature limit \cite{Arnold},
\begin{eqnarray}
 m_h^2(T)
 =
 m_h^2(T=0)
 +
 \left[
  3\lambda_h
  +
  3y_t^2
 \right]\frac{T^2}{12}~,
\end{eqnarray}
where $ y_t $ is the top Yukawa coupling and
$\lambda_h \equiv m_h^2/v^2$ is the quartic coupling of the Higgs boson.
Here, we denote $m_h (T=0)$ by $m_h$.
This limit is valid when the temperature is higher than
the half of masses of particles contributing to the thermal correction, $m < 2T$
\cite{Anderson:1991zb}.
The correction is drastically increased in UED models,
because not only the SM particles but also their KK modes contribute.
We evaluate the contribution from the thermal effect to the masses of Higgs bosons
in the MUED model in this section. 

\begin{figure} [t]
 \begin{center}
 \scalebox{.75}{\includegraphics*{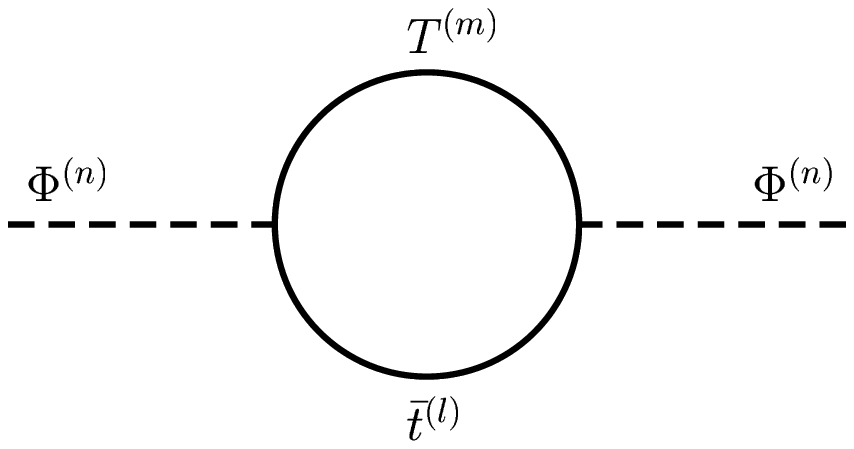}}
 ~~~~~~~~
 \scalebox{.75}{\includegraphics*{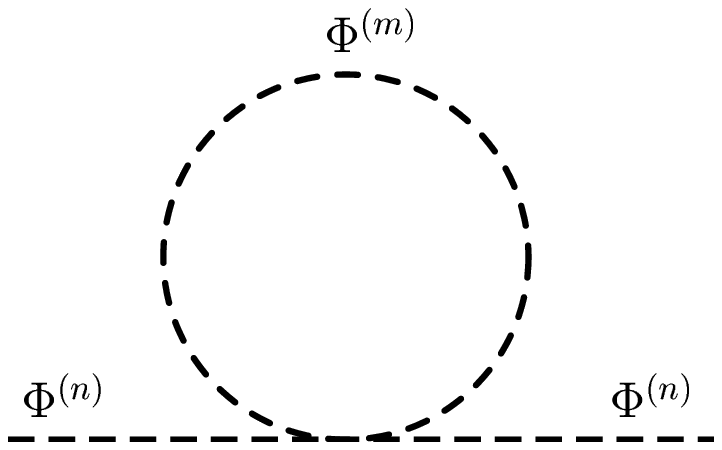}}
 \caption{(Left) Top loop diagram contributing to the thermal mass of Higgs      
          bosons, where $l+m$ or $|l-m|$ is equal to $n$.
          (Right) Higgs loop diagram.}
 \label{fig:thermalloop}
 \end{center}
\end{figure}

As in the case of the SM Higgs boson mentioned above,
the main contribution comes from the top Yukawa interaction and
the self-interaction of the Higgs bosons shown in Fig.~\ref{fig:thermalloop}.
After calculating these diagrams, it turns out that
the thermal mass of the KK Higgs boson $\Phi^{(n)}$ is given by the following formula,
\begin{eqnarray}
 m_{\Phi^{(n)}}^2(T)
 =
 m_{\Phi^{(n)}}^2(T=0)
 +
 \left[
  a (T) \cdot 3\lambda_h
  +
  x (T) \cdot 3 y_t^2
 \right] \frac{T^2}{12}~.
 \label{eq:thermalmass}
\end{eqnarray}
Coefficients $a(T)$ and $x(T)$ are determined by evaluating
how many KK modes contribute to the correction at the temperature $T$.
Particles heavier than $2T$ do not give essential contribution.
Hence, the determination of the coefficients is performed
by counting the number particles lighter than $2T$.

First, let us consider the coefficient $x(T)$.
The masses of KK top quarks, $t^{(n)}$ (left-handed KK top)
and $T^{(n)}$ (right-handed KK top) are approximately $n/R$
before the electroweak symmetry breaking.
Thus, after counting the number of KK top quarks lighter than $2T$,
the coefficient $x(T)$ turns out to be
\begin{eqnarray}
 x(T) =  2[2RT] + 1~,
\end{eqnarray}
where $[x]$ is the integer part of $x$.
It turns out that the $x(T)$ is independent of $n$.
Here, one might think that the masses of the KK top quarks also receive the thermal
correction due to the strong interaction, thus the above counting would be changed.
In fact, both quarks and SU(3) gauge bosons receive the thermal mass from each other
\cite{Weldon,Gatoff}.
However the gauge bosons become very heavy, heavier than $2T$,
in UED models due to the contribution from many (KK) quarks \cite{Weldon}.
As a result, there is no correction of the form $g_s^2T^2$ \cite{Gatoff}
to quark masses, and we can neglect the thermal effect to the top quarks.

Next, we consider the coefficient $a(T)$.
Unlike the contribution from top quarks, the situation is complicated,
because Higgs bosons in the loop diagrams also receive large thermal mass corrections.
In order to evaluate the mass correction correctly,
we employ the resummation method \cite{Arnold}.
According to the method, the thermal correction from Higgs loops,
$\delta m_{\Phi^{(n)}}^2(T)$, satisfies the following consistency equation, 
\begin{eqnarray}
 \delta m_{\Phi^{(n)}}^2(T)
 =
 3\lambda
 \sum_{m = 0}^\infty
 \int\frac{d^3k}{(2\pi)^3} T\sum_{l = 0}^\infty
 \frac{1}{\omega_l^2 + |\vec{k}|^2 +  m_{\Phi^{(n)}}^2(T)}~,
\end{eqnarray}
where $\omega_l = 2\pi i l T$ is the Matsubara frequency.
Substituting expressions for $\delta m_{\Phi^{(n)}}^2(T)$ and $m_{\Phi^{(n)}}^2(T)$
in Eq.~(\ref{eq:thermalmass}) into this equation,
the self-consistency equation for the coefficient $a(T)$ is obtained as 
\begin{eqnarray}
 a(T)
 =
 \sum_{m = 0}^\infty
 \theta
 \left(
  4T^2
  - m^2/R^2
  - x(T) \cdot 3y_t^2\frac{T^2}{12}
  - a(T) \cdot 3\lambda \frac{T^2}{12}
 \right)~.
\end{eqnarray}
Since this equation does not depend on $n$ explicitly,
$a(T)$ is also independent of $n$.
The coefficient $a(T)$ is obtained by solving the equation.
For example, in the case of $m_h =$ 120 GeV, $a(T) =$ 1 for $T < 0.5/R$,
2 for $0.5/R < T < 1/R$, 3 for $1/R < T < 3.5/R$, 2 for $3.5/R < T < 4/R$,
and 0 for $T > 4/R$.
Note that Higgs bosons acquire masses much heavier than $2T$ at the high temperature,
and hence they decouple from the system.
This is automatically taken into account in our calculation of the abundance,
because the number density of a particle is given with a Boltzmann factor.

As can be seen in Eq.~(\ref{eq:thermalmass}),
the Higgs bosons acquire large thermal masses from the thermal bath.
As a result, the Higgs bosons decay into right-handed neutrinos
without a significant phase space suppression.
This fact has a great impact on the production of right-handed neutrinos
in the early universe, which will be discussed in the next section.

\section{Relic abundance of dark matter}

We are now in position to calculate the relic abundance of dark matter
in the UED model with right-handed neutrinos. As
mentioned in the introduction, the thermal history related to the
abundance is given as follows. In the early universe, ($1/R < T < T_R$,
where $T_R$ is the reheating temperature), the first and higher KK
right-handed neutrinos are produced from the thermal bath. Once these
particles are produced, they remain intact in the early universe due to
the very weak Yukawa interaction.  Other KK particles such as the KK
photon have interactions with the thermal bath strong enough to be in
thermal equilibrium. When the temperature of the universe becomes around
$T \sim 1/(20 R)$, the KK photons are decoupled from the thermal bath
(freeze-out).  Finally, in the late universe ($T \ll 1/R$), the first KK
right-handed neutrinos ($N^{(1)}$) are produced from the decays of higher
KK right-handed neutrinos ($N^{(n \geq 2)}$) and KK photons
($\gamma^{(1)}$).  As a result, the relic abundance of the dark
matter is determined by three processes,
\begin{eqnarray}
 \Omega_{\rm DM} 
 =
 \Omega_{\rm thermal}
 +
 \Omega_{N^{(n \geq 2)}{\rm decay}}
 +
 \Omega_{\gamma^{(1)}{\rm decay}}~.
\end{eqnarray}
In this section, we consider these contributions. The parameter region
consistent with the WMAP observation in the framework of the MUED model
with right-handed neutrinos is clarified in the next section.

In most of our quantitative discussions, the cutoff scale is set to be
$\Lambda  = 20/R \sim$ 10 TeV. We assume that the reheating temperature is not
high, at most, $T_R < 20 / R$ for the consistent treatment of UED models.
Hence, the contribution from the KK graviton production
\cite{Feng:2003nr,Shah:2006gs} via thermal scatterings in the early universe is
negligible as shown in Ref. \cite{Shah:2006gs}.

\subsection{Contributions from KK right-handed neutrinos}

First, we consider the relic abundance of the dark matter from the KK
right-handed neutrinos, $\Omega_{\rm thermal} + \Omega_{N^{(n \geq
2)}{\rm decay}}$. As mentioned in Section 2, these KK right-handed
neutrinos are produced through the decay of Higgs bosons. The decay
widths of the Higgs bosons into the KK right-handed neutrino ($N^{(n)}$)
are
\begin{eqnarray}
 \Gamma_{\Phi^{(0)}}^{N^{(n)}}
 &=&
 \frac{y_\nu^2 m_{\Phi^{(0)}}}{8 \pi}
 \left[
  1 - \frac{ (m_{N^{(n)}} - m_{L^{(n)}} )^2 }{ m_{\Phi^{(0)}}^2 }
 \right]
 \beta_f\left(m_{\Phi^{(0)}}, m_{N^{(n)}},m_{L^{(n)}}\right)~,
 \label{eq:dedaywidthzero}
 \\
 \Gamma_{\Phi^{(m)} }^{N^{(n)}}
 &=& 
 \frac{y_\nu^2  m_{\Phi^{(m)}}}{16 \pi}
 \left[
  1 - \frac{ (m_{N^{(n)}} \pm  m_{L^{(|m-n|)}} )^2 }{ m_{\Phi^{(m)}}^2 }
 \right]
 \beta_f \left(m_{\Phi^{(m)}}, m_{N^{(n)}},m_{L^{(|m-n|)}}\right) 
 \nonumber \\
 &+&
 \frac{y_\nu^2 m_{\Phi^{(m)}}}{16 \pi}
 \left[
  1 - \frac{(m_{N^{(n)}} - m_{L^{(m+n)}} )^2}{ m_{\Phi^{(m)}}^2 }
 \right]
 \beta_f\left(m_{\Phi^{(m)}}, m_{N^{(n)}},m_{L^{(m+n)}}\right)~,
 \label{eq:dedaywidthn}
\end{eqnarray}
where the positive (negative) sign is for the case $m-n>0$ $(m-n<0)$.
The function $\beta_f$ comes from the final phase space integration
and is given as
\begin{eqnarray}
 \beta_f (m_i,m_1, m_2)
 \equiv
 \frac{1}{m_i^2}
 \sqrt{m_i^4 - 2 (m_1^2 + m_2^2) m_i^2 + (m_1^2 - m_2^2)^2}~.
\end{eqnarray}
These widths are for the process, $\Phi \rightarrow N\bar{L}$.
Widths for the processes, $ \Phi^* \to \bar{N}L$, are given by the same formulae;
$\Gamma^{\bar N^{(n)}}_{\Phi^{(0,m)*}} = \Gamma^{N^{(n)}}_{\Phi^{(0,m)}}$.

The amount of $N^{(n)}$ produced from the $\Phi^{(m)}$ decay in the early
universe is obtained by solving the following Boltzmann equation,
\begin{eqnarray}
 \frac{d Y^{(n)}}{d T}
 =
 - \frac{ \sum_m C_{(m)}^{(n)} }{ s T H }
 \left(
  1 + \frac{ T }{ 3 g_{*s} (T) } \frac{ d g_{*s} (T) }{ d T }
 \right)~,
 \label{eq:Boltzmann}
\end{eqnarray}
where the entropy density $s$ and the Hubble parameter $H$ are
\begin{eqnarray}
 s = \frac{ 2 \pi^2 }{ 45 } g_{*s} (T) T^3~,
 \qquad
 H = \sqrt{ \frac{ g_{*} (T) }{ 90 } } \frac{\pi  T^2  }{ M_{\rm Pl}  }~.
\end{eqnarray}
Here $ M_{\rm Pl} = 2.4 \times 10^{18} $~GeV is the reduced Planck mass,
and $g_*$ and $g_{*s}$ are the relativistic degrees of freedom of the
thermal bath, which should be treated as the function of the
temperature \cite{Gondolo:1990dk}.
The number density of the KK right-handed neutrinos
($n_{N^{(n)}}$) is normalized by the entropy density as $Y^{(n)} =
n_{N^{(n)}}/s$ in Eq.~(\ref{eq:Boltzmann}). The coefficient
$C_{(m)}^{(n)}$ represents the production rate of $N^{(n)}$ from $
\Phi^{(m)}$, and is defined by
\begin{eqnarray}
 C_{(m)}^{(n)}
 =
 4g_\nu \int \frac{ d^3 k }{ (2 \pi)^3 }
 \gamma_{\Phi^{(m)}}
 \Gamma_{\Phi^{(m)}}^{N^{(n)}} f_{ \Phi^{(m)} }
 \langle 1 - f_L \rangle_{k}~,
\end{eqnarray}
where $\vec{k}$ is the momentum of the Higgs boson, $\gamma_{\Phi^{(m)}}
= m_{\Phi^{(m)}} / (|\vec{k}|^2 + m_{\Phi^{(m)}}^2)^{1/2}$ is the
Lorentz factor, $f_{\Phi^{(m)}} = 1 / (\exp[(|\vec{k}|^2 +
m_{\Phi^{(m)}}^2)^{1/2}/T] - 1)$ is the occupation number of the Higgs
boson, and $\langle 1 - f_L \rangle_k$ is the averaged final-state
multiplicity factor for the final state $L$ with the fixed value of $\vec{k}$.
The factor 4 comes
from the fact that the Higgs field is an SU(2) doublet, and $g_\nu$
stands for the degeneracy of the neutrino mass spectrum; $ g_\nu = 1 $
for the normal hierarchy, $ g_\nu = 2 $ for the inverted hierarchy, and
$ g_\nu = 3 $ for the degenerate case.

The higher KK right-handed neutrinos $N^{(n \geq 2)}$ produce $N^{(1)}$
through cascade decay processes in the late universe\footnote{The lifetime is
short enough for the successful big-bang nucleosynthesis  to be
maintained.}. In order to calculate the contribution $\Omega_{N^{(n \geq
2)}{\rm decay}}$, we have to estimate the number of first KK particles
produced from the decay of one higher KK right-handed neutrino $N^{(n)}$,
which is
denoted by $F^{(n)}$ in the following discussion. Note that the decay of
the KK right-handed neutrino does not conserve the KK number, because KK
number conserving decay are kinematically forbidden. First, trivially
$F^{(1)}=1$, and $N^{(2)}$ cannot produce any first KK particles,
$F^{(2)}=0$. For $ n \geq 3 $, we estimate $F^{(n)}$ in the following
manner.  Suppose that $N^{(n)}$ decays into $D_1^{(n_1)}$ and
$D_2^{(n_2)}$. Here $ n - (n_1 + n_2) = 2m $ where $ m $ is a positive
integer $(m>0)$. We assume that all decay processes occur with the same
probability. For example, $N^{(5)}$ decays into one of the final states
$(n_1,n_2) = (0,3),(1,2),(2,1),(3,0),(0,1),(1,0)$ with the same
probability, $1 / 6$. In the decay of $D_a^{(n_a)}$, both KK number
conserving and violating decays are allowed in general. Therefore, we
assume that the sum of KK number of decay products from $ D_a^{(n_a)} $
are distributed with the same probability. Namely, $n_a, n_a -2, ...,
n_a - k$ have the same branching ratio, $1/k$. Thus, $ D_a^{(n_a)} $
produces $ (n_a + 1) / 2 $ first KK particles for odd $ n_a $ and $ n_a
/2$ first KK particles for even $ n_a $. For example, in the $N^{(5)}$
decay, the number of first KK particles produced through the cascade
decay is $2 \times (2 + 0) / 6 $ from $(n_1,n_2) = (0,3),(3,0)$, $2
\times (1 + 1) / 6 $ from $(n_1,n_2) = (1,2),(2,1)$, and $2 \times (1 +
0) / 6 $ from $(n_1,n_2) = (0,1),(1,0)$. Finally, we obtain the formula
for the fragmentation function $F^{(n)}$ as
\begin{eqnarray}
 F^{(n)}
 = 
 \left\{
 \begin{array}{cl}
  1 & {\rm for}~~ n= 1
  \\
  ( n^2 - 4 )/ ( 3 n ) & {\rm for ~ even}~~ n \geq 2
  \\
  n / 3 & {\rm for ~ odd}~~ n \geq 3 \\
 \end{array}
 \right.~.
\end{eqnarray}

As a result, contributions to the relic abundance of the dark matter
from the KK right-handed neutrinos are
\begin{eqnarray}
 \Omega_{\rm thermal}h^2
 +
 \Omega_{N^{(n \geq 2)}{\rm decay}}h^2
 =
 \frac{ m h^2 s_0}{ \rho_c } \sum_n F^{(n)} Y^{(n)}(T_0)~,
\end{eqnarray}
where $T_0$ is a temperature after the decay of $N^{(n)}$ ends,
$ m = 1/R $ is the $N^{(1)}$ mass, $s_0 = $2889.2 cm$^{-3}$,
$\rho_c =$ 1.053 $\times$ 10$^{-5}$ $h^2$ GeV/cm$^{-3}$, and $h =$
0.73 is the normalized Hubble expansion rate.

It is worth notifying that these contributions are not strongly dependent on the
reheating temperature of the universe. In order to show this result
analytically, we substitute
$\sum_m \Gamma_{\Phi^{(m)}}^{N^{(n)}} n_{\Phi^{(m)}}$ into $C^{(n)}_{(m)}$.
Furthermore, we approximate the number density of the Higgs bosons as
$ n_{\Phi^{(m)}} = 4 \zeta(3) T^3 / \pi^2 $, because they are in thermal
equilibrium in the early universe. Then, the Boltzmann equation 
(\ref{eq:Boltzmann}) can be reduced to
\begin{eqnarray}
 Y^{(n)}
 =
 C \sum_m \int^{T_R}_{T < m_{\Phi^{(m)}}} dT
 \frac{\Gamma_{\Phi^{(m)}}^{N^{(n)}} T^3}{g_{*s} g_*^{1/2} T^6}~
\label{eq:Boltzanna},
\end{eqnarray}
where $C$ is a constant independent of $T$. Here, we ignored the derivative
of $g_{*s}$ to derive this equation. As seen in Eq.~(\ref{eq:Boltzanna}), the production
process is efficient at the low temperature,
and the dependence of the produced abundance on the reheating temperature is weak.
Furthermore, the
abundance is almost independent of $ m_h $, because the thermal correction to
the Higgs boson mass is dominated by top quark loops. Therefore, the
production rate $\Gamma_{\Phi^{(m)}}^{N^{(n)}}$ does not depend on $m_h$.

\subsection{Contribution from KK photon decay}

In order to estimate the relic abundance of the dark matter, we should also
investigate the thermal relic abundance of the KK photon, since the
decay of the KK photon produces the KK right-handed neutrino.
Because the mass difference between the KK photon and the KK right-handed
neutrino is as tiny as $O(0.1)$\% level,
the KK right-handed neutrino abundance produced from the KK photon decay
is just given by the thermal relic abundance of the KK photon, $\Omega_{\gamma^{(1)}}$.

To calculate the thermal relic abundance of the KK photon, we should
consider many first KK particle as coannihilating particles \cite{Kong:2005hn}, because UED
models give a degenerate mass spectrum.  Furthermore, because the mass of
the second KK particles is twice that of the first KK particles,
resonance processes in which second KK particles propagate in the $s$
channel are very important \cite{2ndKK}.  The thermal relic abundance of
the KK photon has been accurately calculated in
Ref.~\cite{Kakizaki:2006dz}, which includes all these effects.
Therefore, we use the result in Ref.~\cite{Kakizaki:2006dz} in order to calculate
the relic abundance of the dark matter from the $\gamma^{(1)}$ decay.

\section{Numerical Results}

In this section, we present the results of the detailed numerical
calculations for the relic abundance of the dark matter in the MUED with
right-handed neutrinos.  In our numerical calculation, the cutoff scale
is set to be $ \Lambda R= 20 $ unless we explicitly show the value of $
\Lambda $.  We have confirmed that all results are independent of $
\Lambda $ as long as $ T_R \lesssim \Lambda / 2 $.  Values of $ g_*$ and
$g_{*s}$ are evaluated by explicit calculation of the contribution from
all KK particles with KK number $ n \le \Lambda$.

\begin{figure} [t]
\begin{center}
 \scalebox{.85}{\includegraphics*{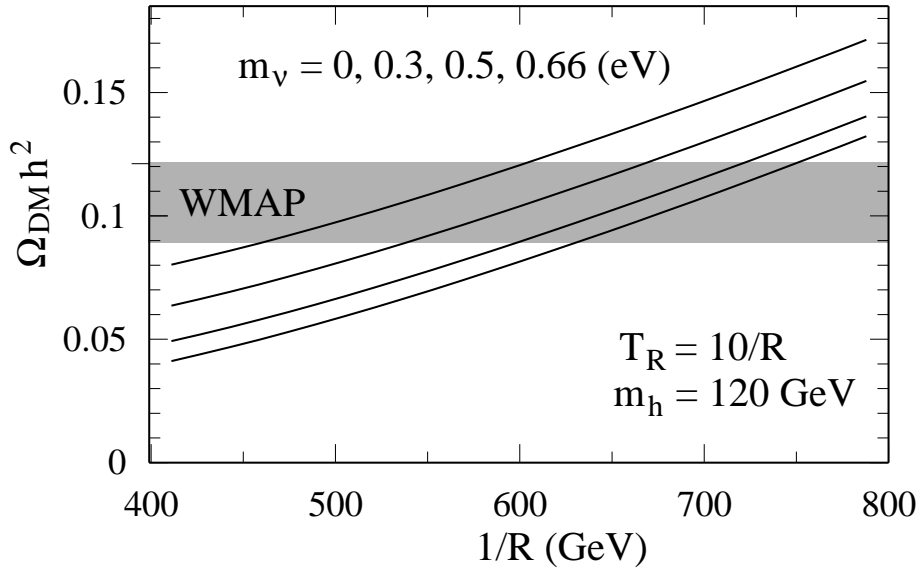}}
 \caption{The dependence of the abundance on $m_\nu$ with fixed $ T_R = 10 /R $ and
          $ m_h = 120 $ GeV. The solid lines are the relic abundance for
		  $ m_\nu = 0, 0.3, 0.5, 0.66 $ eV from bottom to top.  The gray
          band represents the allowed region from the WMAP observation at the
          $2 \sigma$ level}
 \label{fig:mnu}
\end{center}
\end{figure}

The contribution to the abundance from the KK right-handed neutrinos
through the $ \Phi^{(n)} $ decay depends strongly on the decay widths
of the Higgs bosons.  Since all the decay widths are proportional to $
y_\nu^2 $, the abundance is also proportional to $ m_\nu^2 $.  In
Fig. \ref{fig:mnu}, the $m_\nu$-dependence of the abundance is shown
with fixed $ T_R = 10 /R $ and $ m_h = 120 $ GeV.  The solid lines
correspond to the result with $ m_\nu = 0, 0.3, 0.5, 0.66 $ eV from
bottom to top. The first massless neutrino case corresponds to the result
without the contribution from the $ \Phi^{(n)} $ decay, i.e. the MUED case.
The latter three cases correspond to the degenerate mass
spectrum case, i.e. $g_\nu = 3$.  The maximum value of the neutrino
masses, $m_\nu =$ 0.66 eV was obtained by the WMAP observation; $ \sum
m_\nu < 2.0 $~eV \cite{Ichikawa:2004zi}.  The gray band represents the
allowed region from the WMAP observation at the $2 \sigma$ level, $
0.0896 < \Omega_{\rm DM } h^2 < 0.1214 $ \cite{WMAP}.  It can be seen
that the contribution from the KK right-handed neutrinos is negligible
for the normal and inverted hierarchy, $ m_\nu \ll 0.3$ eV. However, for
the maximum neutrino mass, the small compactification scale, $ 1/R
\sim 500 $ GeV, is allowed.
It also can be seen that the contribution from the KK right-handed neutrinos
is almost independent of $1/R$.

\begin{figure} [t]
\begin{center}
\scalebox{.7}{\includegraphics*{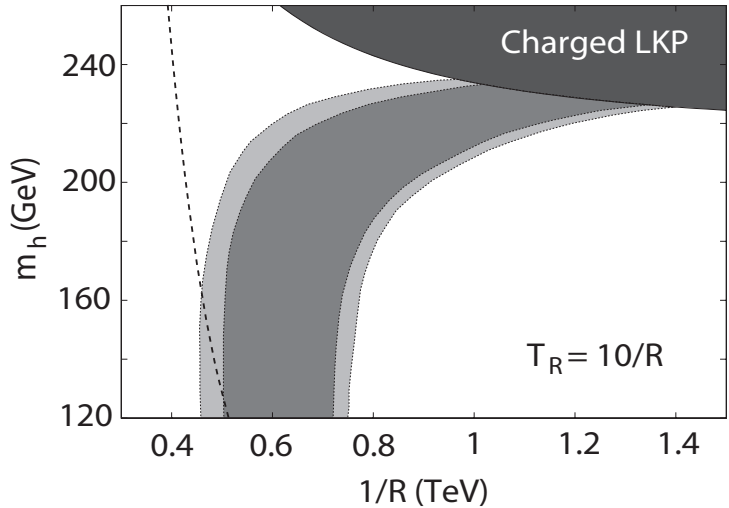}}
\caption{The parameter region of the model consistent with the WMAP
         observation in ($1/R$, $m_h$)-plane. The neutrino mass and the
         reheating temperature are taken as $ m_\nu =$ 0 - 0.66 eV and
         $T_R = 10 / R$. The dark (light) Gray region shows the 1$\sigma$
         (2$\sigma$) allowed region. Dashed line shows the constraint 
         from the electroweak precision measurements \cite{Gogoladze}.
         The right side of dashed line is the allowed region (99\% confidence level).}
\label{fig:mh}
\end{center}
\end{figure}

The parameter region of the model favored by the WMAP observation is
shown in Fig. \ref{fig:mh}.
The result is presented in ($1/R$, $m_h$)-plane with $T_R = 10 /R$.
The neutrino mass is varied from 0 to 0.66 eV.
The dark (light) gray region shows the 1$\sigma$ (2$\sigma$) allowed region\footnote{
For $\Lambda R = 20$, some range of the Higgs mass may be excluded
by additional constraints related to triviality bound ($m_h >$ 200 GeV)
and the vacuum stability ($m_h <$ 150 GeV) \cite{Bhattacharyya:2006ym}.
However, the bound depends on the cutoff scale $\Lambda$,
and the allowed Higgs mass range is significantly expanded for smaller $\Lambda$.
Thus we omitted to draw the bound in the figure.}.
Dashed line shows the constraint 
from the electroweak precision measurements \cite{Gogoladze}. 
The right side of dashed line is the allowed region (99\% confidence level). 
Hence, in this case, some region for small Higgs mass is not favored
by the electroweak precision measurements.
From this figure, it can be seen that
the region is significantly expanded to lower $1/R$ compared to the MUED
model without the right-handed neutrino \cite{Kakizaki:2006dz}.
For $1/R \gtrsim 800$~GeV, the LKP is not the KK graviton but the KK photon,
and the NLKP (next to NLKP) is the KK graviton (KK right-handed neutrinos).
Nevertheless, the estimation of the relic abundance of dark matter
is accomplished by the same procedure.

\begin{figure} [t]
\begin{center}
 \scalebox{.8}{\includegraphics*{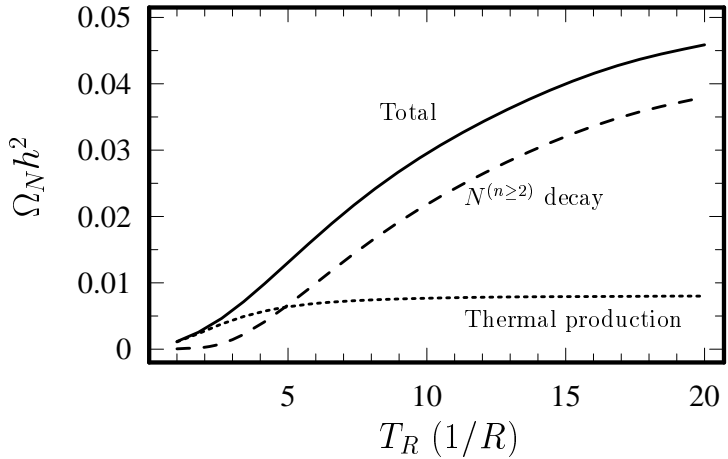}}
 \caption{The dependence of the abundance from the $ \Phi^{(n)} $ decay on $T_R$.
		  The compactification scale, the neutrino mass,
		  and the SM Higgs mass are taken to be
		  $1/R =$ 600 GeV, $m_\nu =$ 0.66 eV, and $ m_h = 120 $ GeV.
          The dotted line represents the abundance of the first right-handed neutrino produced
          from the thermal bath ($\Omega _{\rm thermal}$).
          The dashed line represents the abundance of the the first KK right-handed neutrino produced
          from the decays of higher KK right-handed neutrinos ($\Omega _{N^{(n \geq 2)} {\rm decay}}$).
          The solid line represents the sum of abundances by the $ \Phi^{(n)} $ decay
		($\Omega _{\rm thermal} + \Omega _{N^{(n \geq 2)} {\rm decay}}$). }
 \label{fig:tr}
\end{center}
\end{figure}

Finally, we present the dependence of the abundance
from the $ \Phi^{(n)} $ decay on the reheating temperature.
The result is shown in Fig.~\ref{fig:tr}.
The neutrino and the SM Higgs masses are fixed as
$ m_\nu = $ 0.66 eV and $ m_h =$ 120 GeV in this figure.
The dotted line represents the abundance of the first KK right-handed neutrino produced
from the thermal bath ($\Omega _{\rm thermal}$).
The dashed line represents the abundance of the the first right-handed neutrino produced
from the decays of higher KK right-handed neutrinos ($\Omega _{N^{(n \geq 2)} {\rm decay}}$).
The solid line represents the sum of abundances by the $ \Phi^{(n)} $ decay
		($\Omega _{\rm thermal} + \Omega _{N^{(n \geq 2)} {\rm decay}}$).
The cutoff scale is set to be $ \Lambda R = 50 $.
In order to estimate the dependence on $T_R$,
KK particles  up to $n = 50$ are included in the calculation.
The thermal production of $N^{(1)}$ is saturated soon in higher reheating temperature
as discussed in the previous section.
On the other hand, $\Omega _{N^{(n \geq 2)} {\rm decay}}$ is 
weakly dependent on the reheating temperature,
and the slope becomes gentler for higher reheating temperature.
As a result, for lower reheating temperature $T_R< 5/R$,
$\Omega _{\rm thermal}$ dominates over $\Omega _{N^{(n \geq 2)} {\rm decay}}$
and vice versa.
Hence, the sum of these two contribution is also
weakly dependent on the reheating temperature.
Due to the gentler slope in higher $T_R$,
we can conclude that the uncertainty is not so large
even if we do not know the detail of the reheating process.

\section{Summary}

In this paper, we have investigated the relic abundance of the dark
matter in UED models with right-handed
neutrinos. Due to the small Yukawa interaction, the KK right-handed
neutrinos are out of equilibrium in the early universe. Once they
are produced from the thermal bath, they survive after the freeze-out of
the KK photon and produce the first KK right-handed neutrino in cascade
decay processes.
The contribution from the KK right-handed neutrinos to the relic
abundance of the dark matter is not important for the cases of the
normal and inverted hierarchical neutrino mass spectrum. However, in the
degenerate mass spectrum case, this contribution can be large even if
the reheating temperature is small, $T_R \sim 10/R$. In this case, the
parameter region of UED models consistent with the WMAP observation is
significantly changed.
In the MUED model
with right-handed neutrinos, the compactification scale can be as small
as 500 GeV. This fact has an impact on the collider physics, in
particular on future linear colliders, because first KK particles
can be produced in a pair even if the center of mass energy is around 1
TeV.

\section*{Acknowledgments}

The authors thanks Kenzo Ogure for discussions on the topic of finite temperature field theory.
The work of JS and MS are supported in part by the Grant-in-Aid for the Ministry of
Education, Culture, Sports, Science, and Technology, Government of Japan
(No. 17740131 and 18034001 for JS and No. 18840011 for MS).
The work of MY was financially supported by the Sasakawa Scientific
Research Grant from The Japan Science Society.

\end{document}